\documentclass[manuscript]{acmart}
\AtBeginDocument{%
  }

\usepackage{graphicx}
\usepackage{subcaption}
\acmISBN{}

\begin{document}

\title{Effective Automation to Support the  Human Infrastructure in AI Red Teaming}
\author{Alice Qian Zhang}
\orcid{0009-0005-6407-6981}
\email{aqzhang@andrew.cmu.edu}
\affiliation{%
  \institution{Carnegie Mellon University}
  \city{Pittsburgh}
  \state{Pennsylvania}
  \country{United States}
}

\author{Jina Suh}
\orcid{0000-0002-7646-5563}
\email{jinsuh@microsoft.com}
\affiliation{%
  \institution{Microsoft Research}
  \city{Redmond}
  \state{Washington}
  \country{United States}
}

\author{Mary L. Gray}
\orcid{0000-0001-9972-6829}
\email{mlg@microsoft.com}
\affiliation{%
  \institution{Microsoft Research}
  \city{Cambridge}
  \state{Massachusetts}
  \country{United States}
}

\author{Hong Shen}
\orcid{}
\email{hongs@andrew.cmu.edu}
\affiliation{%
  \institution{Carnegie Mellon University}
  \city{Pittsburgh}
  \state{Pennsylvania}
  \country{United States}
}

\renewcommand{\shortauthors}{Alice Qian Zhang}
\newcommand{\jina}[1]{\textcolor{red}{[Jina: #1]}}

\begin{abstract}

\end{abstract}




\maketitle

Insights summarizing article: 
\begin{itemize}
    \item AI red teaming, like content moderation, relies on human expertise; automation should complement human efforts by enhancing proficiency rather than prioritizing efficiency.
    \item The effectiveness of human expertise in red teaming depends on preserving red teamers' agency, which over-reliance on automation can diminish.
    \item Scaling automation in AI red teaming must account for the necessity of human oversight to ensure adaptive, context-aware risk mitigation.
\end{itemize}

AI systems increasingly make high-stakes decisions, from healthcare diagnostics to financial transactions. Ensuring these technologies align with ethical principles and do not contribute to societal harm is a growing priority. One emerging solution is red teaming--a process that simulates adversarial attacks to uncover vulnerabilities before they can be exploited. 
Red teaming is a well-known practice in national security--originally a vulnerability assessment method for military decision-making that was later adapted for cybersecurity. 
Given these origins, it is no surprise that ``AI'' red teaming gained momentum as an essential practice for AI security and safety mandated by the executive order on ``Safe, Secure, and Trustworthy Development and Use of Artificial Intelligence'' by the Biden Administration in the US.  
Such momentum has been followed by large technology companies, such as OpenAI and Microsoft, to strengthen their red teaming efforts by hiring dedicated professionals, collaborating with domain experts in high-stakes fields such as cybersecurity, finance, and medical, partnering with outsourced contract red teamers through Business Process Outsourcing (BPO) firms, and engaging crowd-sourced gig workers.
At the same time, efforts to broaden red teaming have continued to expand to include open calls for public participation, involving everyday users and impacted communities with lived experiences. 
Examples include the public AI red teaming challenge at the security conference DEFCON 2023, where participants from diverse backgrounds tested AI vulnerabilities. 

As a result, AI red teaming has increasingly relied on a diverse and growing ``human infrastructure''--one that includes hired professionals, outsourced contractors, paid crowdsourced workers and unpaid volunteers, whose labor is critical in identifying vulnerabilities and stress-test contemporary AI systems. 
These red teamers operate along three key dimensions -- payment structure, required expertise, and relationship with the organizing body. At one end of the spectrum, hired domain experts (e.g., security analysts, medical professionals) are often formally employed or contracted for their specialized knowledge, playing a critical role in high-stakes evaluations. Outsourced contractors, engaged through BPOs, are compensated hourly or per task, typically working under structured guidelines. Paid crowdsourced red teamers, such as gig workers on platforms like MTurk and Prolific, perform red teaming tasks on a per-task basis, often with lower initial expertise but the ability to develop domain knowledge over time. Finally, volunteers, including end-users, impacted communities and general public, contribute without compensation, engaging in red teaming out of personal interest or educational opportunities, with varying levels of expertise and an informal relationship with the organizing body.
As AI technologies advance at an unprecedented pace,
supporting such ``human infrastructure'' behind AI red teaming is essential to keeping up with emerging risks and ensuring AI systems remain secure and reliable.

However, the increasing demand for red teaming has also sparked interest in a variety of automated approaches, for example, where AI technologies are used to test other AI systems ~\cite{feffer2024red}. Automation has been explored as a promising solution to improve efficiency, mitigate harm to human red teamers, and enhance scalability. 
The first argument for automation is cost and time efficiency. Scholars argue that, by automating red teaming, organizations may address the breadth of the risks they seek to uncover in a more cost-effective and time-efficient manner~\cite{perez2022red}.
The second argument is risk mitigation for human red teamers. Given the potential harms associated with red teaming such as secondary trauma, well-intentioned efforts exist to develop fully automated approaches that minimize human exposure to distressing or harmful content~\cite{radharapu2023aart}.
The third argument is scalability. As red teaming becomes an essential component of AI safety, there is growing emphasis on scaling up these efforts through automation. The goal is to develop more generalizable and standardized red teaming models that can be broadly applied across different AI systems~\cite{perez2022red, radharapu2023aart}.
  
While automation has the potential to enhance AI red teaming, the way we conceptualize and implement it may inadvertently constrain opportunities for expanding and strengthening the human labor behind red-teaming practices.
Indeed, even with all the calls for automation, current red teaming methodologies lack a standardized definition of how to measure success. 
This inconsistency highlights the need for automation to complement, rather than replace, human judgment in assessing and refining red teaming methodologies.
From our prior work~\cite{zhang2024aura} we see clear parallels between the push to automate AI red teaming and the automation of content moderation, a form of data work that involves identifying and removing harmful content from online platforms.
Below, we discuss how keeping humans in the loop might provide both better red teaming outcomes and safeguard human expertise and well-being using three principles of automation.
We engage critically with emerging views on automated AI red teaming, highlighting opportunities to utilize existing and potential technologies to support human engagement in this work. 
To structure this discussion, we introduce three key pillars of effective automation:
First, we examine the need for \textit{proficiency}, or the development of expertise and skill among red teamers. 
Next, we discuss the importance of \textit{agency} or the ability for red teamers to actively shape, influence, and make meaningful decisions in the process. 
Finally, we illustrate the importance of \textit{adaptability} or the ability of AI red teaming practices to respond to evolving technologies and threats of harm.

\section{The case for proficiency over efficiency}
A key goal of automated AI red teaming is to enhance efficiency and keep pace with the rapid development of AI technologies ~\cite{perez2022red}. 
Existing approaches prioritize maximizing the quantity of red teaming conducted (e.g., generating large datasets of adversarial prompts) while minimizing the cost and time it takes to do so.
These existing approaches view the need for human time and expertise as a fundamental constraint and aim to minimize it. 
The benefits of these advancements are clear: by leveraging automated red teaming, organizations can conduct risk evaluations in a more cost and time-effective manner. However, a singular focus on efficiency risk overlooking how automation might also be leveraged to enhance human expertise and engagement in red teaming.

Content moderation faced a similar challenge when the popularity of social media platforms increased the amount of harmful user-generated content on them. 
As a result, social media platforms began hiring or eliciting volunteer involvement of human personnel to manually review content to determine if it violated platform policies and guidelines. 
However, many large technology companies viewed the cost of maintaining a human moderation workforce as too high, even though content moderators were often underpaid and lacked adequate workplace support for their well-being, as revealed by lawsuits and media reports.
As a result, interest grew in fully automating content moderation to cut costs.
At the same time, automation also promised to increase the volume of content reviewed and speed up the process.
Although many social media platforms use automated methods for content moderation, it remains unclear whether automation improves effectiveness~\cite{gorwa2020algorithmic}.
What has been made clear, however, is that the focus on processing more content at faster speeds has come at a significant cost to moderators.
Initial approaches to automating content moderation potentially contributed to the harsher treatment of moderators by requiring moderators to be more efficient in keeping up without increasing workplace support. 
This has contributed to high workforce turnover rates and explicit backlash from workers who demanded greater workplace support by pursuing class-action lawsuits.

This historical trajectory serves as a cautionary parallel for AI red teaming, where automation is currently being pursued with a similar emphasis on efficiency. If red teaming automation follows the same path, we risk missing critical opportunities to use technology not just to make red teaming efforts more efficient, but also to enhance the skills and expertise of human red teamers, or the \textit{proficiency} of red teamers. 
To support workforce proficiency, we must design tools that support the development of expertise and skill among AI red teamers. We have already surfaced ideas through our prior research from content workers on potential applications of automation ~\cite{zhang2024aura}. Beginning with recruiting new professionals, one promising way is to use automated tools to support the training of new professionals to prepare them for exposure to potentially harmful content. For instance, we may develop training datasets that have graduated levels of severity in exposure.
Additionally, automated tools can support actual work practices such as the automated generation of diverse variations of expert-crafted prompts~\cite{perez2022red}. Ultimately, in this alternative vision of AI red teaming, we argue for the realignment of automation in AI red teaming to amplify the core, essential human aspects of the work, rather than simply treating human involvement as a constraint or cost to be minimized.

\section{The central role of human agency}
Another key motivation behind the development of automated AI red teaming is the desire to reduce human involvement due to the potential risks and harms associated with this work~\cite{radharapu2023aart}.  
For instance, red teaming often requires direct engagement with harmful content, which can lead to psychological distress. 
Additionally, exposure to harmful content can lead to fatigue, and the repetitive nature of tasks like crafting adversarial prompts can exacerbate this effect.
Finally, there may be disproportionate burdens on marginalized workers expected to represent their unique lived experiences.
In this context, automation is positioned as a solution to mitigate these risks by minimizing the direct exposure of human workers to distressing or exploitative content.
However, while the intention to address potential harms to AI red teamers may be well-meaning, it often leads to binary thinking of either completely automated approaches or completely manual processes. Such thinking may limit the consideration of other innovative opportunities for automation.
Even worse, this may take away AI red teamers' ability to provide meaningful oversight on AI red teaming. 

Consider the trajectory that content moderation followed in recent years. 
The initial efforts to spread awareness of the challenges content moderators face--such as harmful working conditions and a lack of workplace support--were crucial. However, framing content moderators primarily as victims led to a push for increased automation, often at the expense of their meaningful contributions, rather than improving their working conditions or leveraging their expertise more effectively. 
For example, even though many of the practitioners we surveyed or interviewed had access to automated content moderation technologies -- such as those designed to limit exposure to harmful content -- some still struggled to integrate these tools into their existing workflows.
For instance, users and content creators have raised concerns that automated moderation systems may wrongly flag appropriate content for review, especially when it covers sensitive topics like LGBTQ+ issues or content from marginalized creators. 
This could significantly increase the workload for content moderators, requiring them to review a high volume of inaccurately flagged content, diverting their attention from genuinely harmful material, and potentially leading to burnout.
This example highlights the need for automation that centers on workers' expertise and ability to make judgment calls -- for example,  we can imagine how this tool could be improved to consider where content moderators' expertise is essential (i.e., accurately identifying violating content that is nuanced) and where automation can provide meaningful assistance (i.e., automatically flagging obvious violations). 

Thus, we propose that more research is conducted to explore automated red teaming approaches that enhance, rather than diminish, the agency of human red teamers. 
Tools for content moderation could be adapted or used as a starting point to explore opportunities to reduce the exposure of workers to harmful content in a way that amplifies parts of red teamer workflows that require their expertise. 
For instance, one might imagine potential automated tools that limit exposure to harmful content without compromising the ability to identify it, 
such as using carbonization or artistic rendering to realistic images that might be otherwise psychologically distressing. 
Ultimately, when we consider automated red teaming as a complement to human red teaming—enhancing rather than replacing human expertise—we uncover numerous opportunities to better support human involvement, ultimately contributing to improved workforce retention and well-being.

\section{The limitations of scaling and standardization}


Finally, much of the justification for automated approaches to AI red teaming hinges on the need to enhance scalability. 
Previous arguments stated that as AI systems grow in complexity and are deployed across increasingly diverse domains, automation enables red teaming efforts to scale by standardizing methodologies. 
This standardization not only makes AI risk assessments more broadly applicable but also ensures consistency in identifying vulnerabilities across different AI models. 
Techniques such as universal filtering mechanisms or generalized adversarial prompts~\cite{feffer2024red} exemplify this shift toward scalable, repeatable red teaming strategies that can be applied across a wide range of AI technologies with minimal adaptation. 
However, this emphasis on scale inherently prioritizes generalizability over adaptability, as scalable methods rely on broad, predefined attack strategies rather than dynamically evolving tactics tailored to specific AI systems. 

Thus, while automation offers clear advantages in terms of consistency and scalability, it also raises concerns about the loss of context-specific \textit{adaptability}. 
For example, some risks that AI red teaming may want to target are deeply embedded in community norms, governance structures, and social dynamics, making certain aspects of red teaming--such as evaluating misinformation, bias, or content moderation--difficult to generalize without losing critical context. 
Again, drawing from our prior work, we found that content moderators often develop domain-specific knowledge over time--such as expertise in identifying instances of violence and terrorism--enabling them to make better-informed decisions about when content should be removed. 
As machine-learning based systems scale, maintaining adaptability becomes increasingly challenging. 
Such large-scale systems often struggle to accommodate local variations in data, models, and deployment environments, leading to rigid, one-size-fits-all solutions that may fail in unpredictable ways~\cite{lwakatare2020large}. 
For example, there are well-documented instances of automated content moderation models removing content that did not violate platform guidelines (i.e., women breastfeeding) or failing to detect content that did violate guidelines (i.e., Facebook failing to detect hate speed during the Myanmar crisis). 
Automated red teaming, while valuable for identifying certain types of technical vulnerabilities, cannot fully replace adaptable, human-expertise-driven practices. 

To address this limitation, red teaming efforts should adopt a hybrid approach, combining scalable automation with targeted, human-driven interventions that remain sensitive to context. In this way, there is much potential to combine approaches of automated generation of prompts ~\cite{perez2022red} with existing human-based approaches. While automation can enhance the scope of AI red teaming, truly effective risk mitigation will require preserving non-scalable interventions, such as case-by-case human oversight and adversarial testing tailored to specific domains.

The future of AI red teaming is at a crossroads. As we have outlined, different visions of automation, labor, and scaling reveal competing priorities. We discussed the tensions of efficiency versus proficiency, the role of human agency, and limitations in scaling red teaming through automation alone. These are not just technical and labor challenges but governance dilemmas that will shape how AI risk mitigation evolves.

If AI red teaming is to remain an effective mechanism for AI safety, it must be carefully designed to navigate these tensions. Automation should not be pursued at the expense of human expertise, nor should the drive to scale red teaming override considerations of adaptability. The true value of automation lies in augmenting human capabilities, not replacing them. The challenge ahead is not merely to automate red-teaming but to ensure that automated systems are designed carefully to empower human experts and support the long-term well-being and retention of the red-teaming workforce. 

Towards this end, we call for researchers, policymakers, and industry practitioners to critically examine the role of automation in AI red teaming and explore how it can serve as an avenue to amplify human expertise rather than diminish it. This means developing tools that support--not supplant--the judgment of red teamers, investing in methodologies that preserve contextual awareness, and fostering career pathways that ensure the long-term well-being and retention of the red-teaming workforce. By centering human expertise within a thoughtfully designed hybrid model, we can build a red teaming ecosystem that is not only scalable but also rigorous, context-aware, and resilient in the face of evolving AI risks.
\bibliographystyle{ACM-Reference-Format}
\bibliography{_references}


\begin{thebibliography}{6}


\ifx \showCODEN    \undefined \def \showCODEN     #1{\unskip}     \fi
\ifx \showDOI      \undefined \def \showDOI       #1{#1}\fi
\ifx \showISBNx    \undefined \def \showISBNx     #1{\unskip}     \fi
\ifx \showISBNxiii \undefined \def \showISBNxiii  #1{\unskip}     \fi
\ifx \showISSN     \undefined \def \showISSN      #1{\unskip}     \fi
\ifx \showLCCN     \undefined \def \showLCCN      #1{\unskip}     \fi
\ifx \shownote     \undefined \def \shownote      #1{#1}          \fi
\ifx \showarticletitle \undefined \def \showarticletitle #1{#1}   \fi
\ifx \showURL      \undefined \def \showURL       {\relax}        \fi
\providecommand\bibfield[2]{#2}
\providecommand\bibinfo[2]{#2}
\providecommand\natexlab[1]{#1}
\providecommand\showeprint[2][]{arXiv:#2}

\bibitem[Feffer et~al\mbox{.}(2024)]%
        {feffer2024red}
\bibfield{author}{\bibinfo{person}{Michael Feffer}, \bibinfo{person}{Anusha Sinha}, \bibinfo{person}{Wesley~H Deng}, \bibinfo{person}{Zachary~C Lipton}, {and} \bibinfo{person}{Hoda Heidari}.} \bibinfo{year}{2024}\natexlab{}.
\newblock \showarticletitle{Red-Teaming for generative AI: Silver bullet or security theater?}. In \bibinfo{booktitle}{\emph{Proceedings of the AAAI/ACM Conference on AI, Ethics, and Society}}, Vol.~\bibinfo{volume}{7}. \bibinfo{pages}{421--437}.
\newblock


\bibitem[Gorwa et~al\mbox{.}(2020)]%
        {gorwa2020algorithmic}
\bibfield{author}{\bibinfo{person}{Robert Gorwa}, \bibinfo{person}{Reuben Binns}, {and} \bibinfo{person}{Christian Katzenbach}.} \bibinfo{year}{2020}\natexlab{}.
\newblock \showarticletitle{Algorithmic content moderation: Technical and political challenges in the automation of platform governance}.
\newblock \bibinfo{journal}{\emph{Big Data \& Society}} \bibinfo{volume}{7}, \bibinfo{number}{1} (\bibinfo{year}{2020}), \bibinfo{pages}{2053951719897945}.
\newblock


\bibitem[Lwakatare et~al\mbox{.}(2020)]%
        {lwakatare2020large}
\bibfield{author}{\bibinfo{person}{Lucy~Ellen Lwakatare}, \bibinfo{person}{Aiswarya Raj}, \bibinfo{person}{Ivica Crnkovic}, \bibinfo{person}{Jan Bosch}, {and} \bibinfo{person}{Helena~Holmstr{\"o}m Olsson}.} \bibinfo{year}{2020}\natexlab{}.
\newblock \showarticletitle{Large-scale machine learning systems in real-world industrial settings: A review of challenges and solutions}.
\newblock \bibinfo{journal}{\emph{Information and software technology}}  \bibinfo{volume}{127} (\bibinfo{year}{2020}), \bibinfo{pages}{106368}.
\newblock


\bibitem[Perez et~al\mbox{.}(2022)]%
        {perez2022red}
\bibfield{author}{\bibinfo{person}{Ethan Perez}, \bibinfo{person}{Saffron Huang}, \bibinfo{person}{Francis Song}, \bibinfo{person}{Trevor Cai}, \bibinfo{person}{Roman Ring}, \bibinfo{person}{John Aslanides}, \bibinfo{person}{Amelia Glaese}, \bibinfo{person}{Nat McAleese}, {and} \bibinfo{person}{Geoffrey Irving}.} \bibinfo{year}{2022}\natexlab{}.
\newblock \showarticletitle{Red teaming language models with language models}.
\newblock \bibinfo{journal}{\emph{arXiv preprint arXiv:2202.03286}} (\bibinfo{year}{2022}).
\newblock


\bibitem[Radharapu et~al\mbox{.}(2023)]%
        {radharapu2023aart}
\bibfield{author}{\bibinfo{person}{Bhaktipriya Radharapu}, \bibinfo{person}{Kevin Robinson}, \bibinfo{person}{Lora Aroyo}, {and} \bibinfo{person}{Preethi Lahoti}.} \bibinfo{year}{2023}\natexlab{}.
\newblock \showarticletitle{Aart: Ai-assisted red-teaming with diverse data generation for new llm-powered applications}.
\newblock \bibinfo{journal}{\emph{arXiv preprint arXiv:2311.08592}} (\bibinfo{year}{2023}).
\newblock


\bibitem[Zhang et~al\mbox{.}(2025)]%
        {zhang2024aura}
\bibfield{author}{\bibinfo{person}{Alice~Qian Zhang}, \bibinfo{person}{Judith Amores}, \bibinfo{person}{Hong Shen}, \bibinfo{person}{Mary Czerwinski}, \bibinfo{person}{Mary~L Gray}, {and} \bibinfo{person}{Jina Suh}.} \bibinfo{year}{2025}\natexlab{}.
\newblock \showarticletitle{AURA: Amplifying Understanding, Resilience, and Awareness for Responsible AI Content Work}.
\newblock \bibinfo{journal}{\emph{Proceedings of the ACM on Human-Computer Interaction}} \bibinfo{volume}{9}, \bibinfo{number}{CSCW2} (\bibinfo{year}{2025}), \bibinfo{pages}{1--45}.
\newblock


\end{thebibliography}

\end{document}